\documentclass[twocolumn,aps,showpacs]{revtex4}

\usepackage{graphicx}

\draft

\begin{document}
\title{Self-trapping of strong electromagnetic beams in relativistic plasmas}
\author{V. I. Berezhiani $^{a,c}$, S.M. Mahajan $^b$, Z. Yoshida$^a$, and M. Ohhashi 
$^a$}
\address{$^a$ Graduate School of Frontier Sciences, The University of Tokyo,\\
Hongo 7-3-1, Tokyo 113-0033, Japan}
\address{$^b$ Institute for Fusion Studies, The University of Texas at Austin, Austin%
\\
78712,Texas, USA}
\address{$^c$ Institute of Physics, Tamarashvili 6, Tbilisi, Georgia}
\date{\today}

\begin{abstract}
Interaction of an intense electromagnetic (EM) beam with hot relativistic
plasma is investigated. It is shown that the thermal pressure brings about a
fundamental change in the dynamics - localized, high amplitude, EM field
structures, not accessible to a cold ( but relativisic) plasma, can now be
formed under well- defined conditions. Examples of the trapping of EM beams
in self-guiding regimes to form stable 2D solitonic structures in a pure e-p
plasma are worked out.
\end{abstract}


\maketitle

\section{Introduction}

The problem of electromagnetic (EM) wave propagation and related phenomena
in relativistic plasmas has attracted considerable attention in the recent
past. From the nonthermal emission of the high-energy radiation coming from
a variety of compact astrophysical objects it has become possible to deduce
the presence of a population of relativistic electrons in the plasma created
in the dense radiation fields of those sources [1]. The principal components
of these plasmas could be either relativistic electrons and nonrelativistic
ions (protons), or relativistic electron-positron (e-p) pairs.

Relativistic e-p dominated plasmas may be created in a variety of
astrophysical situations. Electron-positron pair production cascades are
believed to occur in pulsar magnetospheres [2]. The e-p plasmas are also
likely to be found in the bipolar outflows (jets) in Active Galactic Nuclei
(AGN) [3], and at the center of our own Galaxy [4]. In AGNs, the
observations of superluminal motions are commonly attributed to the
expansion of relativistic e-p beams in a pervading subrelativistic medium.
This model implies copious pair production via $\gamma -\gamma$ interactions
creating an e-p atmosphere around the source. The actual production of e-p
pairs due to photon-photon interactions occurs in the coronas of AGN
accretion disks, which upscatter the soft photons emitted by the accretion
disks by inverse Compton scattering. The presence of e-p plasma is also
argued in the MeV epoch of the early Universe. In the standard cosmological
model, temperatures in the MeV range ($T\sim 10^ {10} K-1 MeV$) prevail up
to times $t =1 s$ after the Big Bang [5]. In this epoch, the main
constituent of the Universe is the relativistic e-p plasma in equilibrium
with photons, neutrinos, antineutrinos, and a minority population of heavier
ions.

Contemporary progress in the development of super strong laser pulses with
intensities $I\sim 10^{21-23}W/cm^{2}$ has also made it possible to create
relativistic plasmas in the laboratory by a host of experimental techniques
[6]. At the focus of an ultrastrong laser pulses, the electrons can acquire
velocities close to the speed of light opening the possibility of simulating
in the laboratory the conditions and phenomena that, generally, belong in
the astrophysical realm [7].

Elucidation of the electromagnetic wave dynamics in a relativistic plasmas
will, perhaps, be an essential tool for understanding the radiation
properties of astrophysical objects as well as of the media exposed to the
field of superstrong laser radiation. Although the study of wave propagation
in relativistic plasmas has been in vogue for some time, it is only in the
recent years that the nonlinear dynamics of EM radiation in e-p dominated
plasmas [8] has come into focus. The enhanced interest stems from two facts
: 1) e-p plasmas seem to be essential constituents of the universe, and 2)
under certain conditions, even an ultrarelativistic electron-proton plasma
can behave akin to an e-p plasma [9].

Wave self-modulation and soliton-formation is, perhaps, one of the more
interesting and significant signatures of the overall plasma dynamics. The
existence of stable localized envelop solitons of EM radiation has been
suggested as a potential mechanism for the production of micropulses in AGN
and pulsars [10]. In the early universe localized solitons are strong
candidates to explain the observed inhomogeneities of the visible universe
[11,12].

In e-p plasmas, there does not exist a general satisfactory theory for the
generation of soliton like structures by ultrastrong high-frequency EM
fields of arbitrary spatio-temporal shape. Relative complexity of the fully
relativistic equations (hydrodynamic or kinetic) has limited their solutions
essentially to one-dimensional systems . A summary of the salient results is
: in unmagnetized e-p plasmas, high-frequency pressure of the EM pulse
pushes e-p pairs in the direction of its propagation thus creating a density
hump in the region of the field localization. The effective refractive index
of medium decreases in this region and as a result localized
soliton-formation is not supported by the medium. (The increase in the
refractive index due to the relativistic nonlinearity related with the
particles' high-frequency motion is not enough to cope with the decrease
caused by the above mentioned effect). In Refs.[11, 13] it has been argued
that localized solitons can be formed if the interaction between the EM
field and acoustic phonons is taken into account- the envelope solitons
propagating with subsonic velocities may, then, emerge. In magnetized e-p
plasma, larger classes of soliton solutions are possible-typical examples
may be found in Refs.[14]. However, it is conceivable that soliton solutions
obtained in a one dimensional formulation will turn out to be unstable in
higher dimensions.

It is, therefore, a matter of utmost priority that we explore the
possibility of finding stable multidimensional soliton solutions in e-p
plasmas. Dynamics of $3D$ envelope solitons of arbitrary strong EM fields in
e-p plasma with a small fraction of heavy ions has been analyzed in
Ref.[12]. It was shown that, in a transparent e-p plasma, EM pulses with $%
L_{||}<<L_{\bot }$ (where $L_{||}$ and $L_{\bot }$ respectively,the
characteristic longitudinal and transverse spatial dimensions of the field),
may propagate as stable, nondiffracting and nondispersing objects (light
bullets) with large density bunching. It was further shown in Ref. [15] that
these bullets are exceptionally robust: they can emerge from a large variety
of initial field distributions, and are remarkably stable. In these studies,
the EM field is pulse-like with longitudinal localization much stronger than
the transverse; the localization is brought about by the charge separation
electric field (usually absent in a pure e-p plasma) created by the presence
of a small population of ions.

In the present paper we explore another mechanism for localization - we will
show that the in the pure e-p plasma, the thermal pressure can provide the
confining ''glue'' just as the charge separation electric field did for an
e-p plasma contaminated with ions. We will also deal with a complimentary
manifestation of the radiation field- the ''beam'' ($L_{\bot }<<L_{||}$)
rather than the pulse ($L_{\bot }>>L_{||}$). Assuming the plasma to be
transparent to the beam, and applying a fully relativistic hydrodynamical
model, we demonstrate the possibility of beam self-trapping leading to the
formation of stable $2D$ solitonic structures. The high-frequency pressure
force of the EM field ( tending to completely expel the pairs radially from
the region of localization) is overwhelmed by the thermal pressure force
which opposes the radial expansion of the plasma creating conditions for the
formation of the stationary self-guiding regime of beam propagation.

\section{Basic Equations}

In this paper the word ''relativistic'' connotes two distinct regimes: the
plasma becomes relativistic when either the directed fluid velocity
approaches the speed of light or the thermal energy per particle is of the
order of, or larger than the rest mass energy. Since both these paths to
relativity are encountered in the astrophysical as well as laboratory
plasmas( produced and accelerated by intense laser pulses), we will
investigate a fully relativistic hydrodynamical model.

If the velocity distribution of the particles of species $\alpha $ ($%
=e,p,i,...$ where $e$,$p$, and $i$ denote respectively electrons, positrons
and heavy ions (protons)) is taken to be a local relativistic Maxwellian,
the hydrodynamics of such fluids is described by [16]: 
\begin{equation}
{\frac{\partial }{\partial x_{k}}}(U_{\alpha }^{i}U_{\alpha k}W_{\alpha })-{%
\frac{\partial }{\partial x_{i}}}P_{\alpha }={\frac{1}{c}}F^{ik}J_{\alpha k}
\end{equation}
where $U_{\alpha }^{i}=[\gamma _{\alpha },\gamma _{\alpha }{\bf u_{\alpha }}%
/c]$ is the hydrodynamic four-velocity with ${\bf u_{\alpha }}$ as the
three-velocity , $\gamma _{\alpha }=(1-u_{\alpha }^{2}/c^{2})^{-1/2}$ is the
relativistic factor, $J_{\alpha k}$ is the four-current, $F^{ik}$ is the
electromagnetic field tensor, and $W_{\alpha }$ is the enthalpy per unit
volume 
\begin{equation}
W_{\alpha }={\frac{n_{\alpha }}{\gamma _{\alpha }}}m_{0\alpha
}c^{2}G_{\alpha }\left( {\frac{T_{\alpha }}{m_{0\alpha }c^{2}}}\right)
\end{equation}
Here $m_{0\alpha }$ and $T_{\alpha }$ are the particle invariant rest mass
and temperature, respectively, $n_{\alpha }$ is the density in the
laboratory frame of the fluid of species $\alpha $. The pressure $P_{\alpha
}=n_{\alpha }T_{\alpha }/\gamma _{\alpha }$, and $G_{\alpha }(z_{\alpha
})=K_{3}(1/z_{\alpha })/K_{2}(1/z_{\alpha })$, where $K_{2}$ and $K_{3}$
are, respectively, modified Bessel functions of the second and third order
and $z_{\alpha }=T_{\alpha }/m_{0\alpha }c^{2}$. The factor $G_{\alpha
}(z_{\alpha })$ has the following asymptotes, $G_{\alpha }\approx
1+5z_{\alpha }/2$ for $z_{\alpha }<<1$ ( non-relativistic) and $G_{\alpha
}\approx 4z_{\alpha }$ for $z_{\alpha }>>1$ ( highly relativistic).

The set of equations (1)-(2) may be written in the standard form : 
\begin{equation}
{\frac{d_{\alpha }}{dt}}(m_{0\alpha }c^{2}G_{\alpha }\gamma _{\alpha })-{%
\frac{1}{n_{\alpha }}}{\frac{\partial P_{\alpha }}{\partial t}}=e_{\alpha }%
{\bf u_{\alpha }}\cdot {\bf E}
\end{equation}
\begin{equation}
{\frac{d_{\alpha }}{dt}}(G_{\alpha }{\bf p_{\alpha }})+{\frac{1}{n_{\alpha }}%
}{\bf \nabla }P_{\alpha }=e_{\alpha }{\bf E}+{\frac{e_{\alpha }}{c}}({\bf %
u_{\alpha }}\times {\bf B})
\end{equation}
where ${\bf p_{\alpha }}=\gamma _{\alpha }m_{0\alpha }{\bf u_{\alpha }}$ is
the hydrodynamical momentum, ${\bf E}$ and ${\bf B}$ are the electric and
magnetic fields, and $d_{\alpha }/dt={\partial /\partial t}+{\bf u_{\alpha }}%
\cdot {\bf \nabla }$ is the comoving derivative. The hydrodynamical velocity 
${\bf u_{\alpha }}$ and the relativistic $\gamma _{\alpha }$ are related to
the momentum by the standard relations: ${\bf u_{\alpha }}={\bf p_{\alpha }}%
/m_{0\alpha }\gamma _{\alpha }$ and $\gamma _{\alpha }=(1+{\bf p_{\alpha }}%
^{2}/m_{0\alpha }^{2}c^{2})^{1/2}$. It is interesting to note from Eqs.
(3)-(4) that the fluid inertia is modified by the temperature; the
expression $M_{eff}^{\alpha }=m_{0\alpha }G_{\alpha }(z_{\alpha })$ denotes
the effective mass of the particle. For ultrarelativistic temperatures ($%
T_{\alpha }>>m_{0\alpha }c^{2}$), the effective mass turns out to be $%
M_{eff}^{\alpha }=4T_{\alpha }/c^{2}>>m_{0\alpha }$. Thus the particles
''forget'' their rest mass and the plasma turns into a kind of ''photon''
gas. If an ultrarelativistic plasma is in thermodynamical equilibrium with
the high-frequency photon gas ($\hbar \omega \sim T$), one should also take
into account the radiation pressure $P_{R}=\sigma T^{4}~~(\sigma =\pi
/45h^{3}c^{3})$ [17]. In this paper this effect will be neglected. We must
also bear in mind that for extremely relativistic regimes, the model
Eqs.(3)-(4) fail to adequately describe the plasma dynamics since heavy
particle production has been neglected. This shortcoming will impose an
upper limit on the temperature for the validity of the model.. Note that in
the context of early universe, the epoch in which the e-p plasma is
dominant, has a characteristic temperature $T_{\alpha }\sim 1MeV$ and $%
M_{eff}\sim 4m_{0e}$. Since the particle masses are just a few times larger
than their rest mass at these temperatures, the e-p plasma can still be
considered as a two component fluid rather than a photon gas.

The equation of state directly follows from the self-consistency of Eqs.(3)
and (4). Taking the scalar product of Eq.(4) with ${\bf u_{\alpha }}$ and
comparing it with Eq.(3), we can derive $d_{\alpha }lnP_{\alpha
}/dt=z_{\alpha }d_{\alpha }G_{\alpha }/dt$. After straightforward
manipulation, the equation can be easily integrated to yield 
\begin{equation}
P_{\alpha }=C{\frac{K_{2}(z_{\alpha })}{z_{\alpha }}}\exp (z_{\alpha
}G_{\alpha })
\end{equation}
where the arbitrary constant $C$ must be defined by the initial state. Using 
$P_{\alpha }=n_{\alpha }T_{\alpha }/\gamma _{\alpha }$ Eq.(5) reduces to the
adiabatic equation of ''state'': 
\begin{equation}
{\frac{n_{\alpha }}{\gamma _{\alpha }}}{\frac{z_{\alpha }}{K_{2}(z_{\alpha })%
}}\exp (-G_{\alpha }z_{\alpha })=const
\end{equation}
For nonrelativistic temperatures, Eq.(6) yields the usual result for a
monoatomic ideal gas ($n_{\alpha }^{r}/T_{\alpha }^{3/2}=const$, where $%
n_{\alpha }^{r}=n_{\alpha }/\gamma _{\alpha }$ is the density in the rest
frame of fluid element) and for ultrarelativistic temperatures one obtains
the equation of state for the photon gas ($n_{\alpha }^{r}/T_{\alpha
}^{3}=const$). We would like to emphasize that the $\gamma _{\alpha }$
factor appearing in Eq.(6) is related to the coherent or directed motion of
fluid elements whose origin may lie either in an initial macroscopic flow or
in the motion imparted by intense EM radiation.

To complete the picture we must couple the plasma equations of motion with
Maxwell equations: 
\begin{equation}
c\nabla \times {\bf B}={\frac{\partial {\bf E}}{\partial t}}+4\pi {\bf J}
\end{equation}
\begin{equation}
c\nabla \times {\bf E}=-{\frac{\partial {\bf B}}{\partial t}}
\end{equation}
\begin{equation}
\nabla \cdot {\bf E}=4\pi \rho
\end{equation}
\begin{equation}
\nabla \cdot {\bf B}=0
\end{equation}
where 
\begin{equation}
\rho =\sum_{\alpha }e_{\alpha }n_{\alpha },~~~~~~{\bf J}=\sum_{\alpha
}e_{\alpha }n_{\alpha }{\bf u_{\alpha }}
\end{equation}
are, respectively, the charge and current densities. The system of
Eqs.(3)-(11) along with the continuity equation ( for each species) 
\begin{equation}
{\frac{\partial n_{\alpha }}{\partial t}}+\nabla \cdot (n_{\alpha }{\bf %
u_{\alpha }})=0,
\end{equation}
represents a closed set of equations which describe propagation of EM
radiation in relativistic multicomponent plasmas.

The above system can be manipulated further to reveal interesting structural
properties. To begin with, Eqs.(5)-(6) can be cast in the form 
\begin{equation}
{\frac{1}{n_{\alpha }}}\nabla P_{\alpha }={\frac{m_{0\alpha }c^{2}}{\gamma
_{\alpha }}}\nabla G_{\alpha }
\end{equation}
which, when substituted into Eq.(4), converts it to 
\begin{equation}
{\frac{\partial }{\partial t}}(G_{\alpha }{\bf p_{\alpha }})+m_{0\alpha
}c^{2}\nabla (G_{\alpha }\gamma _{\alpha })=e_{\alpha }{\bf E}+[{\bf %
u_{\alpha }}\times {\bf \Omega _{\alpha }}]
\end{equation}
where 
\begin{equation}
{\bf \Omega _{\alpha }}={\frac{e_{\alpha }}{c}}{\bf B}+{\bf \nabla }\times
(G_{\alpha }{\bf p_{\alpha }})
\end{equation}
is the so called generalized vorticity . Taking the curl of Eq.(14), we find
that the evolution equation for ${\bf \Omega _{\alpha }}$ 
\begin{equation}
{\frac{\partial {\bf \Omega _{\alpha }}}{\partial t}}={\bf \nabla }\times [%
{\bf u_{\alpha }}\times {\bf \Omega _{\alpha }}]
\end{equation}
is of the standard vortex dynamics form. Although the system of
Eqs.(14)-(16) can be traced to early publications (see for instance
Ref.[18]), their consequences are yet to be fully worked out. An immediate
consequence, for instance, is the appropriate equivalent of Kelvin's
theorem- the flux of generalized vorticity $\Omega _{\alpha }$ is frozen-in
through a comoving area.

The system yields the following set of relativistic Beltrami- Bernoulli
equations for equilibrium [19]: 
\begin{equation}
{\bf \Omega _{\alpha }} = a_{\alpha }{\bf u _{\alpha }}
\end{equation}
\begin{equation}
G_{\alpha }\gamma _{\alpha }+{\frac{e_{\alpha }\phi }{m_{0\alpha }c^{2}}}%
=const
\end{equation}
where $a_{\alpha }$ are constants and $\phi$ is the scalar potential. The
relevance of these equilibria for astrophysics is the subject matter of a
forthcoming paper.

For the current effort, we apply Eqs.(14)-(16) for wave processes in an
unmagnetized plasma. From Eq.(16) it follows that if the generalized
vorticity is initially zero (${\bf \Omega _{\alpha }}=0$) everywhere in
space, it remains zero for all subsequent times. We assume that before the
EM radiation is ''switched on'', the generalized vorticity of the system is
zero. In this case the equation of motion may be written as: 
\begin{equation}
{\frac{\partial }{\partial t}}{\bf \Pi _{\alpha }}+m_{0\alpha }c^{2}\nabla
\Gamma _{\alpha }=e_{\alpha }{\bf E}
\end{equation}
where the temperature dependent momentum ${\bf \Pi _{\alpha }}$ and $\Gamma
_{\alpha }$ are defined by: 
\begin{equation}
{\bf \Pi _{\alpha }}=G_{\alpha }{\bf p_{\alpha }}
\end{equation}
\begin{equation}
\Gamma _{\alpha }=G_{\alpha }\gamma _{\alpha }=\sqrt{G_{\alpha }^{2}+({\bf %
\Pi _{\alpha }}/m_{0\alpha }c)^{2}}
\end{equation}
The condition of vanishing generalized vorticity connects ${\bf \Pi _{\alpha
}}$ with the magnetic field: 
\begin{equation}
{\bf B}=-{\frac{c}{e_{\alpha }}}{\bf \nabla }\times {\bf \Pi _{\alpha }}
\end{equation}
It is remarkable that in Eq.(19) the magnetic part of the Lorentz force is
formally absent; this fact greatly simplifies analytical manipulations. It
is equally remarkable that our equations which describe the dynamics of a
hot relativistic plasma are structurally similar to equations used in the
theoretical treatment of different aspects of ultrastrong laser interaction
with a cold plasma [20]. This similarity becomes even more evident when we
study the interaction of short EM pulse with relativistic electron-ion
plasmas. If the pulse is assumed to be shorter than the characteristic time
for ion response (i.e. inverse of ion Langmuir frequency), the ion motion
may be ignored, and the electric field may be found from the electron part
of Eq.(19), 
\begin{equation}
e{\bf E}=-{\frac{\partial {\bf \Pi _{e}}}{\partial t}}-m_{0e}c^{2}\nabla
\Gamma _{e}
\end{equation}
Substituting this expression into Poisson's Eq.(3) (which now reads as $%
\nabla \cdot {\bf E}=4\pi e(n_{0i}-n_{e})$, where $n_{0i}$ is the
equilibrium ion density) we find the electron density 
\begin{equation}
{\frac{n_{e}}{n_{0i}}}=1+{\frac{1}{m_{0e}\omega _{e}^{2}}}{\frac{\partial }{%
\partial t}}{\bf \Pi _{e}}+{\frac{c^{2}}{\omega _{e}^{2}}}\Delta \Gamma _{e}
\end{equation}
where $\omega _{e}=(4\pi e^{2}n_{0i}/m_{0e})^{1/2}$ is the plasma frequency.
Using Eqs.(22)-(24), Eq.(7) reduces to : 
\[
c^{2}{\bf \nabla }\times {\bf \nabla }\times {\bf \Pi _{e}}+{\frac{\partial
^{2}{\bf \Pi _{e}}}{\partial t^{2}}}+m_{0e}c^{2}{\frac{\partial \nabla
\Gamma _{e}}{\partial t}}+ 
\]
\begin{equation}
+\omega _{e}^{2}{\frac{{\bf \Pi _{e}}}{\Gamma _{e}}}\left[ 1+{\frac{1}{%
m_{0e}\omega _{e}^{2}}}{\frac{\partial }{\partial t}}{\bf \Pi _{e}}+{\frac{%
c^{2}}{\omega _{e}^{2}}}\Delta \Gamma _{e}\right] =0
\end{equation}
which, along with the equation of state ($z_{e}=m_{0e}c^{2}/T_{e}$) 
\begin{equation}
{\frac{n_{\alpha }G_{e}}{\Gamma _{e}}}{\frac{z_{e}}{K_{2}(z_{e})}}\exp
(-G_{e}z_{e})=const,
\end{equation}
constitutes the simplified system to which the entire set of Maxwell and
relativistic hydrodynamic equations for an the electron -ion plasma have
been reduced. An equation similar to Eq.(25) was derived in the cold plasma
limit in Ref.[21]. However, there are important differences : a) Due to the
temperature dependence of $G_{e}$ in Eq.(21), the factor $\Gamma _{e}$ and
the momentum $\Pi _{e}$ are no more related by simple relations as they do
for a ''cold'' case, and b) to incorporate the temperature variation in the
system we must add the equation of state (26).

Though Eqs.(25)-(26) form a more complicated system than what we have for
the cold plasma, we believe that many results obtained in the cold plasma
limit can find appropriate analogies in the hot relativistic- temperature
case. Detailed studies of the nonlinear dynamics of the electron-ion plasma
is beyond the intended scope of the current paper and some new results will
be presented separately elsewhere. In the next part of the current paper, we
concentrate on the nonlinear dynamics of EM beams in e-p dominated plasmas.

\section{The electron-positron dominated plasma}

In this section we apply our general formulation to the problem of
self-trapping of EM beams in pure e-p plasmas with relativistic
temperatures. For notational convenience, we replace the subscripts (e) and
(p) by superscripts $(-)$, and $(+)$. We assume that the equilibrium state
of the plasma is characterized by an overall charge neutrality $n_{\infty
}^{-}=n_{\infty }^{+}=n_{\infty }$, where $n_{\infty }^{-}$ and $n_{\infty
}^{+}$ are the unperturbed number densities of the electrons and positrons
in the far region of the EM beam localization. In most mechanisms for
creating e-p plasmas, the pairs appear simultaneously and due to the
symmetry of the problem it is natural to assume that $T_{\infty
}^{-}=T_{\infty }^{+}=T_{\infty }$, where $T_{\infty }^{-}$ and $T_{\infty
}^{+}$ are the respective equilibrium temperatures .

We shall assume that for the radiation field of interest, the plasma is
underdense and transparent, i.e., $\epsilon =\omega _{e}/\omega \ll 1$,
where $\omega $ is the mean frequency of EM radiation and $\omega _{e}=(4\pi
e^{2}n_{\infty }/m_{0e})^{1/2}$ is the plasma frequency. Since both species
are mobile, the e-p dynamics can not be reduced to just one vector equation
similar to Eq.(25). We will display the entire set in terms of potentials (
the Coulomb gauge ${\bf \nabla }\cdot {\bf A}=0$ will be used), 
\begin{equation}
{\bf E}=-{\frac{1}{c}}{\frac{\partial {\bf A}}{\partial t}}-{\bf \nabla }%
\phi ,~~~~~~~{\bf B}={\bf \nabla }\times {\bf A},
\end{equation}
and the dimensionless quantities ${\tilde{t}}=\omega t$, ${\tilde{{\bf r}}}%
=(\omega /c){\bf r}$, $\tilde{T}^{\pm }=T^{\pm }/m_{0e}c^{2}$, ${\tilde{{\bf %
A}}}=e{\bf A}/(m_{0e}c^{2})$, ${\tilde{{\phi }}}=e{\phi }/m_{0e}c^{2}$, $%
\tilde{{\bf \Pi }}^{\pm }={\bf \Pi }^{\pm }/(m_{0e}c)$, and $\tilde{n}^{\pm
}=n^{\pm }/n_{\infty }$. Suppressing the label- ''tilde'', we may arrive at
the dimensionless equations, 
\begin{equation}
{\frac{\partial {\bf \Pi }^{\pm }}{\partial t}}+{\bf \nabla }\Gamma ^{\pm
}=\mp {\frac{\partial {\bf A}}{\partial t}}\mp {\bf \nabla }\phi
\end{equation}

\begin{equation}
{\frac{\partial ^{2}{\bf A}}{\partial t^{2}}}-\Delta {\bf A}+{\frac{\partial 
}{\partial t}}{\bf \nabla }\phi -\epsilon ^{2}({\bf J^{+}}-{\bf J^{-}})=0
\end{equation}
\begin{equation}
\Delta \phi =\epsilon ^{2}(n^{-}-n^{+})
\end{equation}
\begin{equation}
{\bf \nabla }\cdot {\bf A}=0
\end{equation}
\begin{equation}
{\frac{\partial n^{\pm }}{\partial t}}+{\bf \nabla }{\bf J}^{\pm }=0
\end{equation}
with ${\bf J}^{\pm }=n^{\pm }{\bf \Pi }^{\pm }/\Gamma ^{\pm }$ and $\Gamma
^{\pm }=\sqrt{(G^{\pm })^{2}+({\bf \Pi }^{\pm })^{2}}$. The species equation
of state is: 
\begin{equation}
{\frac{n^{\pm }}{\Gamma ^{\pm }f(T^{\pm })}}={\frac{1}{\Gamma _{\infty
}^{\pm }f(T_{\infty })}}
\end{equation}
where 
\begin{equation}
f(T^{\pm })={\frac{K_{2}(1/T^{\pm })T^{\pm }}{G^{\pm }}}\exp [G^{\pm
}/T^{\pm }]
\end{equation}
Of various techniques that could be invoked to investigate Eqs.(28)-(34) to
study the self-trapping of high-frequency EM radiation propagating along the 
$z$- axis, we choose the method presented in the excellent paper by Sun et
al. [22]. The method is based on the multiple scale expansion of the
equations in the small parameter $\epsilon $. Assuming that all variations
are slow compared to the variation in $\xi =z-at$, we expand all quantities $%
Q=({\bf A},\phi ,{\bf \Pi }^{\pm },n^{\pm },...)$ as 
\begin{equation}
Q=Q_{0}(\xi ,x_{1},y_{1},z_{2})+\epsilon Q_{1}(\xi ,x_{1},y_{1},z_{2})
\end{equation}
where $(x_{1},y_{1},z_{2})=(\epsilon x,\epsilon y,\epsilon ^{2}z)$ denote
the directions of slow change, and $a_{1}=(a^{2}-1)/\epsilon ^{2}\sim 1$. We
further assume that the high-frequency EM field is circularly polarized, 
\begin{equation}
{\bf A_{0\bot }}={\frac{1}{2}}({\bf {\hat{x}}}+i{\bf {\hat{y}}})A\exp (i\xi
/a)+c.c.
\end{equation}
Here $A$ is the slowly varying envelope of the EM beam, ${\bf {\hat{x}}}$
and ${\bf {\hat{y}}}$ denote unit vectors, and $c.c.$ is the complex
conjugate. We now give a short summary of the steps in the standard
multiple- scale methodology (Ref.[22]). To the lowest order in $\epsilon $,
we obtain the following. The transverse (to the direction of EM wave
propagation $z$) component of Eq.(28) reduces to 
\begin{equation}
{\bf \Pi _{0\bot }^{\pm }}=\mp {\bf A_{0\bot }}
\end{equation}
and for the longitudinal components we get: 
\begin{equation}
-a{\frac{\partial \Pi _{0z}^{\pm }}{\partial \xi }}+{\frac{\partial \Gamma
_{0}^{\pm }}{\partial \xi }}={\mp }(-a){\frac{\partial A_{0z}}{\partial \xi }%
}{\mp }{\frac{\partial \phi _{0}}{\partial \xi }}.
\end{equation}
Equations (29)-(31) yield $\partial _{\xi }{\bf \nabla _{\bot }}\phi
_{0}=\partial _{\xi }^{2}\phi _{0}=\partial _{\xi }A_{z0}=0$, where ${\bf %
\nabla }_{\bot }$ is the perpendicular Laplacian in $(x_{1},y_{1})$. These
relations imply that $\phi $ and $A_{0z}$ do not depend on the fast variable 
$\xi $. For the self-trapping problem, we can assume that $A_{0z}=\Pi
_{0z}^{\pm }=0$ [22]. From Eq.(38), and from the lowest order continuity Eq.
(32), we obtain : $\partial _{\xi }\Gamma _{0}^{\pm }=\partial _{\xi
}n_{0}^{\pm }=0$, i.e., $\Gamma _{0}^{\pm }$ and $n_{0}^{\pm }$ also do not
depend on the fast variable $\xi $.

To the next order (in $\epsilon $), the transverse component of Eq.(28)
reads : 
\begin{equation}
-a{\frac{\partial {\bf \Pi _{1\bot }}^{\pm }}{\partial \xi }}+{\bf \nabla }%
_{\bot }\Gamma _{0}^{\pm }={\mp }(-a){\frac{\partial {\bf A_{1\bot }}}{%
\partial \xi }}\mp {\bf \nabla _{\bot }}\phi _{0}
\end{equation}
Averaging over the fast variable $\xi $ we obtain ${\bf \nabla _{\bot }}%
\Gamma _{0}^{\pm }=\mp {\bf \nabla _{\bot }}\phi _{0}$ yielding the trivial
solution $\phi _{0}=0$ and $\Gamma _{0}^{\pm }=\Gamma _{0}=const$. Note that
from Eqs.(30) and (33), we can deduce that $n_{0}^{+}=n_{0}^{-}=n_{0}$ and $%
T_{0}^{+}=T_{0}^{-}=T_{0}$.

Thus, as one would expect, the low frequency motion of the e-p plasma is
driven by the ponderomotive pressure ($\sim {\bf \Pi _{0\bot }}^{2}$) of the
high-frequency EM field and this force, being same for the electrons and
positrons, does not cause charge separation. It is also evident that due to
the symmetry between the electron and positron fluids, their temperatures,
being initially equal, will remain equal during the evolution of the system.
The relation between the EM field and the temperature can be found using the
equation $\Gamma _{0}=const$ obtained above. Using Eqs.(36)-(37) and by
choosing the $const$ by requiring that at infinity $A\to 0$ and $T_{0}\to
T_{\infty }$, we derive 
\begin{equation}
G^{2}(T_{0})=G^{2}(T_{\infty })-|A|^{2}
\end{equation}
It follows from Eq.(40) that the present hydrodynamical model, which
describes the nonlinear waves in e-p plasma, is valid for $|A|^{2}/G_{\infty
}^{2}\le 1$. In our opinion the origin of this restriction lies in the
inadequacy of the basic model, and is not solely due to a failure of the
perturbation technique used above. When this condition is violated, the EM
waves are overturned and they will cause multistream motion of the plasma
(i.e. wave breaking takes place). In such a situation, one must resort to
kinetic description for studying the nonlinear wave motion. Notice however
that the function $G(T_{\infty })\to 1$ if $T_{\infty }\to 0$ but rapidly
increases with increase of $T_{\infty }$ thus providing room for $|A|_{\max
} $ to reach from weak to relativistic values.

We are now ready to deal with the equation for the slowly varying envelope $%
A $ of the EM beam. To the lowest order in $\epsilon $, one finds from
Eq.(29) 
\begin{equation}
a_{1}{\frac{\partial ^{2}{\bf A_{0\bot }}}{\partial \xi ^{2}}}-{\bf \nabla }%
_{\bot }^{2}{\bf A_{0\bot }}-2{\frac{\partial ^{2}{\bf A_{0\bot }}}{\partial
\xi \partial z_{2}}}+2{\frac{n_{0}(T_{0})}{G_{\infty }}}{\bf A_{0\bot }}=0
\end{equation}
For deriving this equation, we used the relation $\Gamma _{0}=\sqrt{%
G^{2}(T_{0})+|A|^{2}}=G_{\infty }$, and 
\begin{equation}
n_{0}(T_{0})={\frac{f(T_{0})}{f(T_{\infty })}}
\end{equation}

which follows from Eq.(33). Substituting Eq.(36) into Eq.(41) we find: 
\begin{equation}
2i{\frac{\partial A}{\partial z}}+{\bf \nabla }_{\bot }^{2}A+{\frac{2}{%
G_{\infty }}}[1-n(T_{0})]A=0
\end{equation}
where subscripts for variables ($z_{2},x_{1},y_{1}$) are dropped for
simplicity. We also assumed without loss of generality that $%
(a^{2}-1)/\epsilon ^{2}a^{2}=2/G_{\infty }$, which in dimensional units
coincides with the linear dispersion relation of the EM wave in an e-p
plasma, namely: $\omega ^{2}=2\omega _{e}^{2}/G_{\infty }+k^{2}c^{2}$
provided that $a=\omega /kc$.

Thus, the dynamics of EM beams in hot relativistic e-p plasma has become
accessible within the context of a generalized nonlinear Schr\"{o}dinger
equation (NSE) (43).

\section{The self-trapped beams in e-p plasma}

In this section we seek the localized 2D soliton solutions of Eq.(43), and
analyze the stability of such solutions. Making the self-evident
re-normalization of variables $z\to zG_{\infty }$ , $r_{\bot }\to r_{\bot }%
\sqrt{G_{\infty }/2}$, Eq.(43) can be written as: 
\begin{equation}
i{\frac{\partial A}{\partial z}}+{\bf \nabla }_{\bot }^{2}A+\Psi A=0
\end{equation}
where $\Psi =1-n_{0}(T_{0})$ represents the generalized nonlinearity. The
companion equation (40) can be viewed as a transcendental algebraic relation
between $T_{0}$ and $|A|^{2}$. Thus we conclude that $\Psi $ is a function
of $|A|^{2}$ ($\Psi =\Psi (|A|^{2})$. We note that Eq.(44) can be written in
the Hamiltonian form $iA_{z}=\delta H/\delta A^{*}$, where $H=\int d{\bf %
r_{\bot }}[|\nabla _{\bot }A|^{2}-F(|A|^{2})]$ and $F(t)=\int_{0}^{t}\Psi
(t^{^{\prime }})dt^{^{\prime }}$. The Hamiltonian structure implies that
Eq.(44) conserves the Hamiltonian $H$ in addition to the power (''photon
number'') $N=\int d{\bf r_{\bot }}|A|^{2}$.

Unfortunately, it is not possible, in general, to derive an explicit
analytic relation $\Psi =\Psi (|A|^{2})$ for arbitrary value of $T_{\infty }$%
. Some qualitative deductions readily follow. Equation (40) shows that the
presence of EM radiation reduces the temperature $T_{0}$. Since $%
df(T_{0})/dT_{0}>0$, from Eq.(42) we conclude that the plasma density is
also reduced in the region of the EM field localization which is in
accordance with adiabatic motion of the plasma. For higher strength of the
EM field, a complete expulsion of plasma i.e. plasma cavitation can take
place ($n_{0}\to 0$); this has been predicted in Ref.[23]. Thus the
nonlinearity function $\Psi $ shows a saturating character with the increase
of EM field strength (note that present model is valid provided $%
|A|^{2}/(G_{\infty }^{2}-1)\le 1$). To illustrate, we exhibit in Fig.1 a
plot of $\Psi $ versus $|A|^{2}$ for $T_{\infty }=0.1$. One can see that the
nonlinearity function indeed saturates at high intensity.
\begin{figure} \begin{center}
	\includegraphics[width=7cm]{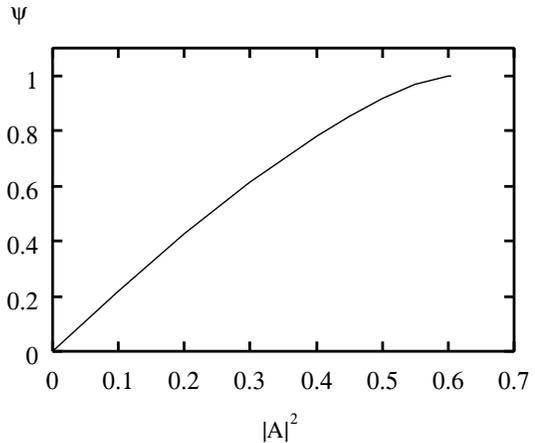}
\end{center}
\caption{The nonlinearity function $\Psi $ versus $|A|^{2}$ for $%
T_{\infty }=0.1$.}
\end{figure}
For small
temperatures, we can even obtain an analytic expression for the function $%
\Psi $. Remembering $T_{0}\le T_{\infty }$, assuming $T_{\infty }\ll 1$, and
by using Eq.(42) along with the asymptotic expansions $G_{0}(\approx
1+5T_{0}/2)$ and $f(\approx T_{0}^{3})$, we derive for the nonlinearity
function: 
\begin{equation}
\Psi =1-\left( 1-{\frac{|A|^{2}}{5T_{\infty }}}\right) ^{3/2}
\end{equation}
Equations (44)-(45) admit a stationary, nondiffracting axially symmetric
solution of the form ${A/\sqrt{5T_{\infty }}}=U(r)\exp (i\lambda z)$ where $%
r=(x^{2}+y^{2})^{1/2}$ and $\lambda $ is the nonlinear wave-vector shift.
The radially dependent envelope $U(r)$ obeys an ordinary nonlinear
differential equation: 
\begin{equation}
{\frac{d^{2}U}{dr^{2}}}+{\frac{1}{r}}{\frac{dU}{dr}}-\lambda U+\Psi
(U^{2})U=0
\end{equation}
where $\Psi =1-(1-U^{2})^{3/2}$. This equation corresponds to a boundary
value problem with the boundary conditions: $U$ has its maximum $U_{m}$ at $%
r=0$, and $U\to 0$ as $r\to \infty $. We remind the reader that it has been
shown in a seminal paper of Vakhitov and Kolokolov [24] that such solutions
exist for arbitrary saturating nonlinearity functions $\Psi $, provided that
the eigenvalue $\lambda $ satisfies $0<\lambda <\Psi _{m}$, where $\Psi _{m}$
is a maximal value of the nonlinearity function. Equation (46) admits an
infinity of discrete bound states characterized by $j=0,1,2...$ zeros at
finite $r$. We consider only the lowest-order nodeless solution of Eq.(46),
i. e. ''ground state'' that is positive and monotonically decreasing with
increasing $r$. In the asymptotic region the solution must decay as $U_{r\to 
{\infty }}\sim \exp (-{\sqrt{\lambda }}r)/\sqrt{\lambda r}$. Our
nonlinearity function $\Psi $ has a maximum $\Psi _{m}=1$ found at $%
U_{m}(=1) $, i. e. at the maximally allowed strength of the field. As a
consequence the upper bound of the propagation constant $\lambda _{c}$ must
satisfy $\lambda _{c}<\Psi _{m}$. Numerical simulations show that the
amplitude of the ground state solution $U_{m}=U(r=0,\lambda )$ is a growing
function of $\lambda $ (see Fig.2) and it acquires its maximum value ($=1$)
at $\lambda =\lambda _{c}\approx 0.29$.
\begin{figure} \begin{center}
	\includegraphics[width=7cm]{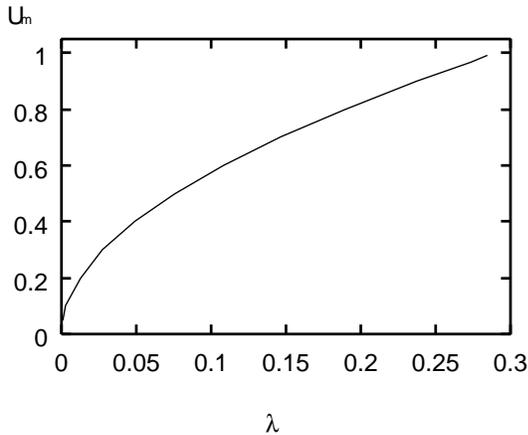}
\end{center}
\caption{Nonlinear dispersion relation: the amplitude $U_{m}$ as a
function of $\lambda $, the eigenvalue. The plasma temperature $T_{\infty
} << 1$.}
\end{figure}
The solution represents a trapped,
localized EM solitary beam. The beam becomes wider at low amplitudes.

The stability of the solitonic solutions can be investigated using the
criterion of Vakhitov and Kolokolov [24]- the soliton is stable against
small, arbitrary perturbations if $dN/d\lambda >0$, where $N$ is the power
of the trapped mode: 
\begin{equation}
N(\lambda )=2\pi \int_{0}^{\infty }dr~rU^{2}(r,\lambda )
\end{equation}
\begin{figure} \begin{center}
	\includegraphics[width=7cm]{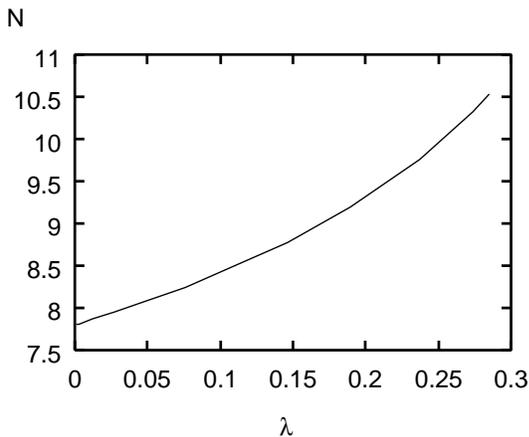}
\end{center}
\caption{The beam power $N$ versus $\lambda $ ($T_{\infty }<<1$).}
\end{figure}
In Fig.3 we plot the numerically obtained solutions of $N$ for various $%
\lambda $. Since the curve has positive slope everywhere, the corresponding
ground state solution is stable for $0<\lambda <\lambda _{c}$. Notice that
the power of the solitary beam always exceeds a certain critical value $%
N>N_{c}\approx 7.8$ . We also know that $N$ must be bounded from above ($%
N\le N_{m}\approx 10.5$).

For arbitrary temperatures, explicit form of $\Psi =\Psi (|A|^{2})$ can not
be found. However, due to its saturating character, solutions with
properties similar to the small temperature case (which can be explicitly
solved) could be expected. Using relations (34), (40) and (42), we
numerically find a stationary solution of Eq.(44) for arbitrary $T_{\infty }$%
. For convenience we use following representation of vector potential ${%
A/A_{c}=U(r)\exp (i\lambda z)}$, where $A_{c}=(G_{\infty }^{2}-1)^{1/2}$.
Though the maximum value of $U$ is still restricted by the condition $%
0<U_{m}\le 1$, the amplitude of vector potential $A_{m}$ can reach a
considerable value. For ultrarelativistic temperatures, $T_{\infty }>>1$, we
have $A_{c}=\sqrt{15T_{\infty }}>>1$ and since $0<A_{m}\le A_{c}$ the
soliton solution with ultrarelativistic strength of EM field is possible.
Here we present results of simulations for $T_{\infty }=1$ (i.e. $T_{\infty
}\approx 0.5MeV$ ).
\begin{figure} \begin{center}
	\includegraphics[width=7cm]{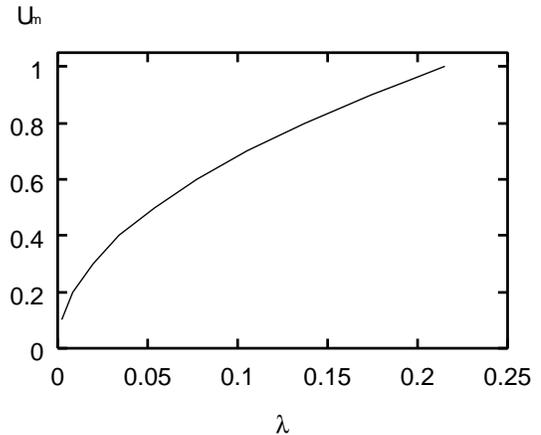}
\end{center}
\caption{The nonlinear dispersion relation, $U_{m}$ versus $\lambda $
for $T_{\infty }=1$.}
\end{figure}
In Fig.4 we plot the amplitude of the ground state
solution $U_{m}$ versus the propagation constant $\lambda $. The solution
exists provided $0<\lambda <\lambda _{c}(\approx 0.22)$. The profiles of the
field $U(r)$ the plasma density $n_{0}(r)$ and the temperature $T_{0}(r)$
are exhibited in Fig.5 for $\lambda =0.1$.
\begin{figure} \begin{center}
	\includegraphics[width=7cm]{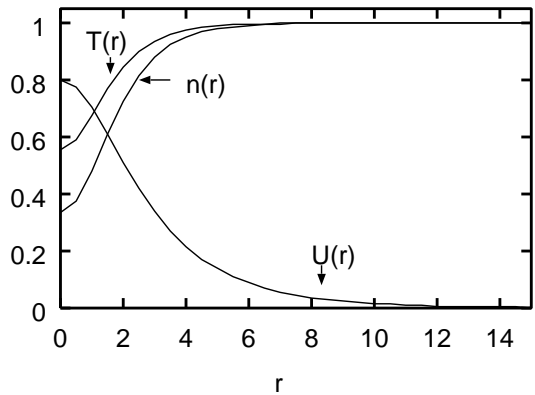}
\end{center}
\caption{Normalized EM field $U$, plasma temperature $T_{0}$ and
density $n_{0}$ versus $r$ for $T_{\infty }=1$.}
\end{figure}
One can see that in the region of
field localization, the plasma temperature and density is reduced. Similar
plots could be obtained for all allowed values of $\lambda $. When $\lambda
\to \lambda _{c}$, plasma cavitation takes place, i.e. at $r=0$ the plasma
density and temperature tends down to zero. Appearance of zero temperature
is not surprising since the corresponding region is the ''plasma vacuum'';
all particles are gone away.

\begin{figure} \begin{center}
	\includegraphics[width=7cm]{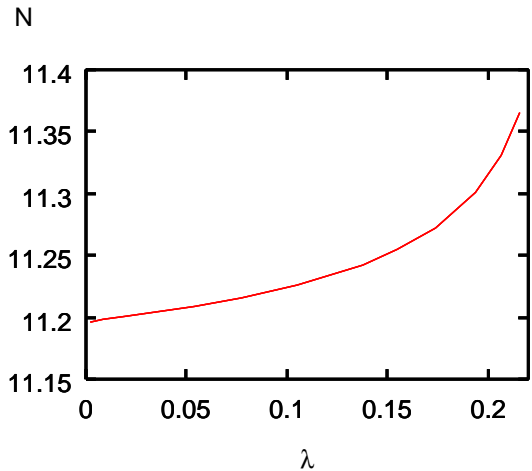}
\end{center}
\caption{The beam power $N$ versus $\lambda $ ($T_{\infty }=1$).}
\end{figure}
The dependence of $N$ on $\lambda $ is presented in Fig.6. One can see that
the curve $N=N(\lambda )$ has a positive slope and according to Vakhitov and
Kolokolov criterion, the corresponding solitary solutions are stable against
small perturbations.

The detailed dynamics of arbitrary field distribution must be studied by
direct simulations of Eq.(44). We can learn much , however, from the recent
extensive elucidations of the dynamical properties of the solutions of NSE
with saturating nonlinearity. It seems that the general features of
evolution are not sensitive to the details of the saturating nonlinearity
(see for instance [15, 25] and references therein). For all such systems the
beam will monotonically diffract if the beam power is below a critical value
($N<N_{c}$), and it will be trapped if $N>N_{c}$ and the Hamiltonian $H<0$
is negative. In the latter case, the beam parameters will oscillate near the
equilibrium, ground state values. This oscillations monotonically decrease
with increase $z$ due to the appearance of the radiation spectrum. For
larger $z$, the oscillations are damped out, and the formation of the
soliton in its ground state takes place. If the initial profile of the beam
is close to the equilibrium one, the beam quickly reaches the ground-state
equilibrium, and propagates for a long distance without distortion of its
shape. The initial beam, even when its parameters (i.e. amplitude, effective
width and phase) are quite far from equilibrium, will either focus or
defocus to the ground state, exhibiting damped oscillations around it. As a
consequence the beam reaches an equilibrium with its final power slightly
smaller than the initial. Such an evolutionary scenario may not hold for
very intense beams with $N>>N_{c}$; the beam may then break up into
filaments due to a modulation instability. However, each filament, will tend
to evolve towards its own equilibrium state corresponding to the power it
carries. Thus, the ground-state equilibrium seems to be an attractor.

Our own studies indicate that Eq.(44), with the nonlinearity particular to
the problem at hand, reproduces the general expected behavior described
above. However we find that the soliton formation requires the initial beam
power to be in the range $N_{c}<N<N_{m}$. For $N>N_{m}$, the multistream
motion of the plasma prevents the system from settling in a steady state.

\section{conclusions}

We have investigated the nonlinear propagation of strong EM radiation in a
relativistic, unmagnetized two-fluid plasma. The treatment is fully
relativistic- in the coherent or directed motion as well as in the random or
thermal motion of the plasma particles. The assumption that prior to the
switching of the field-plasma interaction, the generalized vorticity is
zero, greatly simplifies the system of relativistic fluid equations. In
particular, in the electron-ion dominated plasma, under well defined
conditions the system of Maxwell -fluid equation ( Eqs.(25)-(26)) turns out
to be structurally similar to the one obtained for a cold plasma.
Consequently we would expect that results already established in cold plasma
limit can find appropriate analogy in the hot plasma case.

We presented a somewhat detailed study of EM beam propagation in transparent
e-p plasmas. Applying a reductive perturbation technique, the system of
relativistic Maxwell- fluid equations is reduced to a 2D nonlinear
Schr\"{o}dinger equation with a saturating nonlinearity. We found that if
the strength of the EM field amplitude is below the wave breaking limit, the
beam can enter the self-trapped regime resulting in the formation of stable,
self-guided $2D$ solitonic structures. The beam-trapping owes its origin to
the thermal pressure (which opposes the ponderomotive pressure) - Naturally
such trapping regimes are not accessible in the relativistic but cold plasma
limit. In the region of beam trapping, the plasma density as well as its
temperature is reduced and under certain conditions these parameters can be
reduced considerably (i.e. plasma cavitation takes place).

The fact that relativistically hot e-p plasmas are capable of sustaining
high amplitude localized structures of high amplitude electromagnetic fields
should be a result of considerable importance to an understanding of the
complex radiative properties of different astrophysical objects where such
plasmas are known to exist.

The work of Z Y is partially supported by Toray Science Foundation, the work
of V I B was partially supported by the INTAS Georgian CALL-97, and S M M's
work was supported by the U.S. Department of Energy Contract No.
DE-FG03-96ER-54346.



\end{document}